\documentclass[preprint,showpacs,preprintnumbers,amsmath,amssymb,
superscriptaddress,floatfix]{revtex4}
\usepackage{graphicx}
\usepackage{amsmath}
\usepackage{bm}
\usepackage{mathrsfs}
\usepackage{color}
\usepackage{slashed}
\usepackage{dcolumn}

\newcommand{\be}{\begin{equation}}
\newcommand{\ee}{\end{equation}}
\newcommand{\ba}{\begin{eqnarray}}
\newcommand{\ea}{\end{eqnarray}}
\newcommand{\nn}{\nonumber}

\newcommand{\la}{\langle}
\newcommand{\ra}{\rangle}

\begin{document}


\title{$D \bar{D}^*$ and $\pi \psi $ interactions in a unitary
coupled-channel approximation }

\author{Bao-Xi Sun}
\email{sunbx@bjut.edu.cn}
\affiliation{College of Applied Sciences, Beijing University
of Technology, Beijing 100124, China}
\affiliation{Department of Physics, Peking University,
Beijing 100871, China}

\author{Da-Ming Wan}
\affiliation{College of Applied Sciences, Beijing University
of Technology, Beijing 100124, China}

\author{Guang-Yi Tang}
\affiliation{Institute of High Energy Physics, Chinese Academia of Sciences, Beijing 100049, China}

\author{Fang-Yong Dong}
\affiliation{College of Applied Sciences, Beijing University
of Technology, Beijing 100124, China}

\date{\today}

\begin{abstract}
The $D \bar{D}^*$ interaction via a $\pi \psi$ intermediate state is studied carefully in the isospin $I=1$ sector. By solving the Bethe-Salpeter equation in the unitary coupled-channel approximation, we obtain the S-wave amplitude as a function of the total energy of the system in the center of mass frame. A resonance state is generated dynamically in the 3900MeV region, which might correspond to the $Zc(3900)$ particle. Moreover, the loop function of a vector meson and a pseudoscalar meson is deduced explicitly in the dimensional regularization scheme and the contribution of the longitudinal part of the vector meson propagator is taken into account. The initial and final polarization vectors in the vertex of the vector meson and the pseudoscalar meson are eliminated when the Bethe-Salpeter equation is solved, and it is certified that the amplitude is still unitary in the calculation.
\end{abstract}


\maketitle

\section{Introduction}

%






In the past decade many exotic particles with hidden heavy-quark flavors have been observed experimentally, and these observations stimulate people's interests in studying the properties and structures of these exotic particles theoretically. More information on the experimental and theoretical research works on this topic can be found in the review articles of Refs.~\cite{Yuan_CZ,Chen_HX,Pilloni:review}.
In 2013, the BESIII Collaboration studied the $e^+ e^- \rightarrow J/\psi \pi^+ \pi^-$  process, and observed a peak distribution in the $J/\psi \pi^\pm$ invariant mass spectrum\cite{Ablikim:2013mio}, The mass and decay width of this particle take the values of $M=3899.0\pm3.6\pm4.9MeV$ and $\Gamma=46\pm10\pm20MeV$, respectively.
A later analysis on the $(D \bar{D}^*)^\pm$ invariant mass spectrum in the $e^+ e^- \rightarrow \pi^\pm (D \bar{D}^*)^\mp$ process supplied a resonance state with mass $3883.9\pm4.5$MeV and width $24.8\pm11.5$MeV, and the quantum number of this state is determined as $I^G(J^{PC})=1^+(1^{+-})$ with the angular distribution analysis\cite{Ablikim:2013xfr}. In 2015, a neutral structure near the $D \bar{D}^*$ threshold was observed in the processes of $e^+ e^- \rightarrow \pi^0 \pi^0 J/\psi$ and $e^+ e^- \rightarrow (D \bar{D}^*)^0 \pi^0$\cite{Ablikim:2015a,Ablikim:2015b}.

The charmonium-like state $Zc(3900)$, observed by BESIII Collaboration in the process of $e^+ e^- \rightarrow J/\psi \pi^+ \pi^-$\cite{Ablikim:2013mio} and then confirmed by Belle and CLEO Collaborations through the same process\cite{Liu:2013dau,Xiao:2013iha},
has inspired more discussions theoretically.
Initially $Zc(3900)$ is assumed to be a tentra-quark state, which consists of a $\bar{c}c$ and a light quark-antiquark pair
\cite{Voloshin:2013dpa,Wang:2013vex,Faccini:2013lda,Terasaki:2013lta,Dias:2013xfa,Qiao:2013raa,Braaten:2013boa,
Navarra}.
Since the $Zc(3900)$ particle is close to the $D \bar{D}^*$ threshold, and it is naturally to be regarded as a
$D \bar{D}^*$ molecule state\cite{Sun:2012zzd,Zhang:2013aoa,Cui:2013yva,Zhao:2014gqa,Chen:2015ata,Dong:2013iqa,Guo:2013sya,Wang:2013cya,
Wilbring:2013cha,He:2015mja,Guo:2013ufa,Zhou:2015jta,Patel:2014vua,Albaladejo:2015lob,Prelovsek:2013xba,
Chen:2014afa,Aceti:2014uea,Gong:2016hlt}.
Furthermore, some people think that these structures might come from some kinematical effects, such as the triangle singularities\cite{
 Szczepaniak:2015eza, Liu:2015taa}, and the coupled channel cusp effect\cite{Swanson:2014tra,Chen:2013coa, Liu:2013vfa,Ikeda:2016zwx}.
In Ref.~\cite{Pilloni:1612}, the different scenarios are analysed and it is concluded that the current data is not precise enough to distinguish between these hypotheses.

In this work, the interaction Lagrangian of $D \bar{D}^*$ and $\pi \psi$ in Ref.~\cite{Gong:2016hlt} is adopted,  and then the vertices for the processes of $D \bar{D}^* \rightarrow D \bar{D}^*$ and $D \bar{D}^* \rightarrow \pi \psi$ are obtained. In the unitary coupled-channel approximation, the Bethe-Salpeter equation is solved with a loop function where the longitudinal part of the vector meson propagator is taken into account. A resonance state near the $D \bar{D}^*$ threshold is generated dynamically, and it is assumed to be associated to the $Zc(3900)$ particle observed experimentally.



This article is organized as follows. The theoretical framework is described in Section~\ref{sect:framework}. The calculation results are presented in Section~\ref{sect:results}. A summary is given in Section~\ref{sect:summary}, and the derivation of the loop function formula related to the longitudinal part of the vector meson propagator is  presented in the appendix part.

\section{Theoretical framework}
\label{sect:framework}

the contact $D \bar{D}^*$ four-point interaction lagrangian takes the following form when the heavy quark symmetry
is considered\cite{Gong:2016hlt},
\be
\label{eq:contact}
{\sl L}=\lambda_1 \la ( D \bar{D}^{* \mu} + h. c. )^2 \ra,
\ee
where the field operators $D$ and $D^*$ are the $SU(2)$ isospin doublets,
\be
D~=~
\left(\begin{array}{c}
                                               D^+  \\
                                               D^0
                                               \end{array}
                                          \right),
~~~~
D^*~=~
\left(\begin{array}{c}
                                               D^{*+}  \\
                                               D^{*0}
                                               \end{array}
                                          \right),
\ee
and the symbol $\la ... \ra$ denotes the trace in the $SU(2)$ isospin space.

The interaction Lagrangian related to the $D$, $D^*$ $J/\psi$ and $\pi$ particles can be written as
\ba
\label{eq:DDpsipi}
{\sl L}_{D D^* \psi \pi}&=& \left. \lambda_2 \nabla_\nu \psi_\mu \la \bar{D}^{*\mu} u^\nu D \ra
                          + \lambda_3 \psi_\mu \la \nabla^\nu \bar{D}^{* \mu} u_\nu D \ra \right. \nn \\
                        &+& \left. \lambda_4 \nabla_\nu \psi_\mu \la \bar{D}^{* \nu} u^\mu D \ra
                          + \lambda_5 \psi_\mu \la \nabla^\mu \bar{D}^{* \nu} u_\nu D  \ra + h.c. \right.,
\ea
where
$\psi^\mu$ stands for $J/\psi$, $\nabla_\mu$ is a covariant derivative operator,
and $u_\mu=i(u^\dagger \partial_\mu u - u \partial_\mu u^\dagger)$ with
\be
u=\exp \left( \frac{i\phi}{\sqrt{2}f_\pi} \right),~~~~\phi=\left(\begin{array}{cc}
                                            \frac{\pi^0}{\sqrt{2}}   &  \pi^+  \\
                                              \pi^- & -\frac{\pi^0}{\sqrt{2}}
                                               \end{array}
                                          \right),
\ee
  and the pion decay constant $f_\pi=93$MeV.

The interaction potential between the $D$ and $\bar{D}^*$ mesons can be obtained from the lagrangian in
Eq.~(\ref{eq:contact}),
\be
\label{eq:VDD}
V_{D \bar{D}^* \rightarrow D \bar{D}^*}= \lambda_1 C_{ij} \varepsilon \cdot \varepsilon^*,
\ee
with $\varepsilon$ and $\varepsilon^*$ the polarization vectors of the initial and final
vector mesons, respectively.
The coefficients $C_{ij}$ in the different channels are depicted in Table~\ref{table:coef_DDstar}

\begin{table}[htbp]
\begin{tabular}{c|cccc}
\hline\hline
 $C_{ij}$          &$D^+ D^{*-}$ & $D^0 \bar{D}^{*0}$ & $\bar{D}^{0} D^{*0}$ & $D^- D^{*+}$    \\
\hline
 $D^+ D^{*-}$      &$-2$ & $-2$ & $-2$ & $-4$    \\
 $D^0 \bar{D}^{*0}$&$$   & $-2$ & $-4$ & $-2$    \\

$\bar{D}^{0}D^{*0}$&$$ & $$   & $-2$ & $-2$    \\

$D^- D^{*+}$       &$$   & $$   & $$   & $-2$    \\
 \hline \hline
\end{tabular}
\caption{The coefficients $C_{ij}$ in the $D$ and $D^*$ meson interaction, $C_{ji}=C_{ij}$.}
\label{table:coef_DDstar}
\end{table}


According to the isospin, parity and $C$-parity of the $Zc(3900)$ particle, we can construct a $D \bar{D}^*$ pair  with isospin $I=1$
\be
| D \bar{D}^*, I=1 \ra = \frac{1}{\sqrt{2}} \left(|D^+ D^{*-} \ra - |D^0 \bar{D}^{*0}\ra
-|\bar{D}^{0} D^{*0}\ra + |D^- D^{*+}\ra \right),
\ee
where the $C$-parity is negative and the usual assignment for $D\sim i \bar{q} \gamma_5 c$ and $D^{*}_{\mu} \sim \bar{q} \gamma_\mu c$, with $C D C^{-1}=D^{ \dagger}$ and $C D^{ *}_\mu C^{-1}=-D^{ *\dagger}_\mu$.

Thus the potential of $D \bar{D}^* \rightarrow D \bar{D}^* $ in the isospin $I=1$ sector takes the form of
\be
\label{eq:VDDDD}
V_{D \bar{D}^* \rightarrow D \bar{D}^*}= -8 \lambda_1 \varepsilon \cdot \varepsilon^*.
\ee

The interaction potential for the process of $D^+ D^{*-} \rightarrow \pi^0 \psi$ is deduced from the Lagrangian in
Eq.~(\ref{eq:DDpsipi}),
\ba
\label{eq:Vddpsipi0}
V_{D^+ D^{*-} \rightarrow \pi^0 \psi}=\frac{1}{f_\pi}
\left[ - \lambda_2(p_2 \cdot k_2 ) \varepsilon \cdot \varepsilon^*
+ \lambda_3 (p_1 \cdot k_2 ) \varepsilon \cdot \varepsilon^* \right.
- \left. \lambda_4 (p_2 \cdot \varepsilon ) (k_2 \cdot \varepsilon^* )
+ \lambda_5 (p_1 \cdot \varepsilon^* ) (k_2 \cdot \varepsilon ) \right],
\ea
where $k_1$ and $k_2$ are the momenta of the initial and final pseudoscalar mesons, and
$p_1$ and $p_2$ are those of the initial and final vector mesons, respectively.
Actually, since the zeroth component of the polarization vectors tends to zero as the three-momentum of the particles goes to zero, the third and fourth terms in Eq.~(\ref{eq:Vddpsipi0}) indeed can be neglected for small kinetic energies of the particles. Therefore, only the first and second terms are taken into account in the following discussion and calculation.

Similarly, the potential for the process of $D^- D^{*+} \rightarrow \pi^0 \psi$ is denoted as
\be
V_{D^- D^{*+} \rightarrow \pi^0 \psi}=V_{D^+ D^{*-} \rightarrow \pi^0 \psi}.
\ee

The potential in the processes of $D^0 \bar{D}^{*0} \rightarrow \pi^0 \psi$ and $\bar{D}^0 {D}^{*0} \rightarrow \pi^0 \psi$ both take the negative value of the potential
of $D^+ D^{*-} \rightarrow \pi^0 \psi$, i.e.,
\be
V_{D^0 \bar{D}^{*0} \rightarrow \pi^0 \psi}=-V_{D^+ D^{*-} \rightarrow \pi^0 \psi},
\ee
and
\be
V_{\bar{D}^0 {D}^{*0} \rightarrow \pi^0 \psi}=-V_{D^+ D^{*-} \rightarrow \pi^0 \psi}.
\ee

Therefore, in the isospin $I=1$ sector, the potential of $D \bar{D}^* \rightarrow \pi^0 \psi$
can be written as

\ba
\label{eq:VDDpsipi}
V_{D \bar{D}^* \rightarrow \pi^0 \psi}=
{2\sqrt{2}}V_{D^+ D^{*-} \rightarrow \pi^0 \psi}.
\ea
It is apparent that the potential of $\pi^0 \psi \rightarrow D \bar{D}^*$ takes the same form as that
in Eq.~(\ref{eq:VDDpsipi}).


Actually, the potentials in Eqs.~(\ref{eq:VDDDD}) and ~(\ref{eq:VDDpsipi}) only supply interaction vertices when the Bethe-Salpeter equation is solved, while the out-lines related to the initial and final vector mesons in the  Feynmann diagrams should be cut off. It means that the polarization vectors of the initial and final vector mesons, $\varepsilon$ and $\varepsilon^*$, in Eqs.~(\ref{eq:VDDDD}) and ~(\ref{eq:VDDpsipi})
should be eliminated when we try to solve the Bethe-Salpeter equation.
%

The contact potential of $D \bar{D}^*$ in Eq.~(\ref{eq:VDDDD}) can be written as
\be
\label{eq:VDDDD-2}
V_{D \bar{D}^* \rightarrow D \bar{D}^*}= \tilde{V}_{D \bar{D}^* \rightarrow D \bar{D}^*} g^{\mu \nu} \varepsilon_\mu  \varepsilon^*_\nu,
\ee
where
\be
\label{eq:tildeV1}
\tilde{V}_{D \bar{D}^* \rightarrow D \bar{D}^*}=-8 \lambda_1.
\ee
Similarly, the potential for the process of $D \bar{D}^* \rightarrow \pi^0 \psi$ in Eq.~(\ref{eq:VDDpsipi}) is denoted as

\be
\label{eq:VDDpsipi-2}
V_{D \bar{D}^* \rightarrow \pi^0 \psi} = \tilde{V}_{D \bar{D}^* \rightarrow \pi^0 \psi}g^{\mu \nu} ~\varepsilon_\mu~\varepsilon_\nu^*,
\ee
where
\be
\label{eq:tildeV2}
\tilde{V}_{D \bar{D}^* \rightarrow \pi^0 \psi}=\frac{2\sqrt{2}}{f_\pi}
\left[ - \lambda_2 (p_2 \cdot k_2 )
+ \lambda_3 (p_1 \cdot k_2 )
\right],
\ee
with $p_2 \cdot k_2=\frac{s-M_\psi^2-m_\pi^2}{2}$ and $p_1 \cdot k_2=\frac{u-M_{\bar{D}^*}^2-m_\pi^2}{-2}$. The
Mandelstam variables $s=(p_2+k_2)^2$ and $u=(p_2-k_1)^2$, and in the heavy meson approximation, $u\approx (p^0_2-k^0_1)^2$.

In the dimensional regularization, the loop-function in the Bethe-Salpeter equation take the following form
 \ba
 \label{eq:Gpr}
G_{ab}(s)&=&i \int \frac{d^4 q}{(2\pi)^4} \frac{1}{q^2-M_a^2+i\epsilon } \frac{1}{(P-q)^2-M_b^2+i\epsilon} \nn \\
&=& \frac{1}{16 \pi^2} \left\{ a_l(\mu) + \ln \frac{M_a^2}{\mu^2}
+ \frac{M_b^2-M_a^2 + s}{2s} \ln \frac{M_b^2}{M_a^2} + \right.
\nonumber \\ & &  \phantom{\frac{2 M_a}{16 \pi^2}} +
\frac{\bar{q}_l}{\sqrt{s}} \left[
\ln(s-(M_a^2-M_b^2)+2\bar{q}_l\sqrt{s})+
\ln(s+(M_a^2-M_b^2)+2\bar{q}_l\sqrt{s}) \right.   \\
& & \left. \phantom{\frac{2 M_a}{16 \pi^2} +
\frac{\bar{q}_l}{\sqrt{s}}} \left. \hspace*{-0.3cm}-
\ln(-s+(M_a^2-M_b^2)+2\bar{q}_l\sqrt{s})-
\ln(-s-(M_a^2-M_b^2)+2\bar{q}_l\sqrt{s}) \right] \right\}, \nn \ea
with the square of the total energy of the system $s=P^2$ and the three-momentum of the intermediate particles in the center of mass frame
\begin{equation}
\bar{q}_l=\frac{\sqrt{s-(M_a+M_b)^2}\sqrt{s-(M_a-M_b)^2}}{2\sqrt{s}}.
\end{equation}

The loop-function in Eq.~(\ref{eq:Gpr}) is used to study the pseudoscalar meson - vector meson system\cite{Nagahiro:2008cv}, the vector-vector meson system in the unitary coupled-channel approximation\cite{Molina:2008jw, Geng:2008gx}. Moreover, after the on-shell approximation is considered, this formula is also used in the calculation of the vector meson-baryon scattering amplitude\cite{Oset:2009vf, Gonzalez:2008pv, Sarkar09}. However, the contribution from the longitudinal part of the vector meson propagator is not considered in the loop function in Eq.~(\ref{eq:Gpr}) when the pseudoscalar meson - vector meson interaction is discussed. Here we will take into account the longitudinal propagator of the vector meson, and then the pseudoscalar meson - vector meson loop function in the dimensional regularization scheme takes the form of
\ba
G_l(s)&=&i\int \frac{d^4 q}{(2\pi)^4} \frac{-g_{\mu \nu}+\frac{q_\mu q_\nu}{M_a^2}}{q^2-M_a^2+i\varepsilon}
\frac{1}{(P-q)^2-M_b^2+i\varepsilon} \nn \\
&=&-g_{\mu \nu} \left[ G_{ab}(s)+\frac{1}{M_a^2}H^{00}_{ab}(s) \right]-\frac{P_\mu P_\nu}{M_a^2} H^{11}_{ab}(s),
\ea
and the meanings of $H^{00}_{ab}(s)$ and $H^{11}_{ab}(s)$ can be found in the appendix part. Apparently, the loop function $G_l(s)$ can be rewritten as
\be
\label{eq:loop-vector-pseudo}
G_l(s)= g_{\mu \nu} \tilde{G}_l(s),
\ee
with
\be
\label{eq:20170820}
\tilde{G}_l(s)=-\left( G_{ab}(s)+\frac{1}{M_a^2}H^{00}_{ab}(s)+\frac{s}{4M_a^2}H^{11}_{ab}(s)\right).
\ee
Clearly the terms including $H^{00}_{ab}(s)$ and $H^{11}_{ab}(s)$ in the loop function in Eq.~(\ref{eq:20170820}) are related to the longitudinal part of the vector meson propagator, which is not taken into account in the previous works.

If the potentials in Eqs.~(\ref{eq:VDDDD-2}) and~ (\ref{eq:VDDpsipi-2}) and the loop function in Eq.~(\ref{eq:loop-vector-pseudo}) are substituted into the Bethe-Salpeter equation, we would obtain
\be
\tilde{T}g^{\mu \nu}=\tilde{V}g^{\mu \nu}+\tilde{V}g^{\mu \alpha}~g_{\alpha \beta}\tilde{G}~\tilde{V}g^{\beta \nu}+..., \label{eq:Bethe}
\ee
and thus
\ba
\tilde{T}&=&\tilde{V}+\tilde{V}\tilde{G}\tilde{V}+... \nn \\
&=&[1-\tilde{V}\tilde{G}]^{-1} \tilde{V}. \label{eq:Bethe}
\ea
The amplitude $\tilde{T}$ is unitary when the Bethe-Salpeter equation is solved.

In Ref.~\cite{Sarkar09}, where the interaction of the vector meson and the baryon decuplet is studied, it is  assumed that $\varepsilon \cdot \varepsilon^*=-3$,
%
%
while the value of $\varepsilon \cdot \varepsilon^*$ is set to be $-1$ in Refs.~\cite{Abreu:2011ic,Altenbuchinger:2013vwa,Sun:2016tmz}. Anyway, all these assumptions are reasonable.
However, in Ref.~\cite{Sun2014}, the polarization vectors $\varepsilon$ and $\varepsilon^\prime$ in the potential of the vector meson and the baryon octet are replaced by their matrix forms, and $\varepsilon \cdot \varepsilon^*$ becomes a function of the scattering angle. Now it must be emphasized that the treatment in Ref.~\cite{Sun2014} is not correct and it results in the resonance peaks generated dynamically are all close to the real axis in the complex plane of the total energy $\sqrt{s}$ in the center of mass frame.

\section{Results}
\label{sect:results}

\begin{figure}[!htb]
\centerline{
\includegraphics[width = 0.65\linewidth]{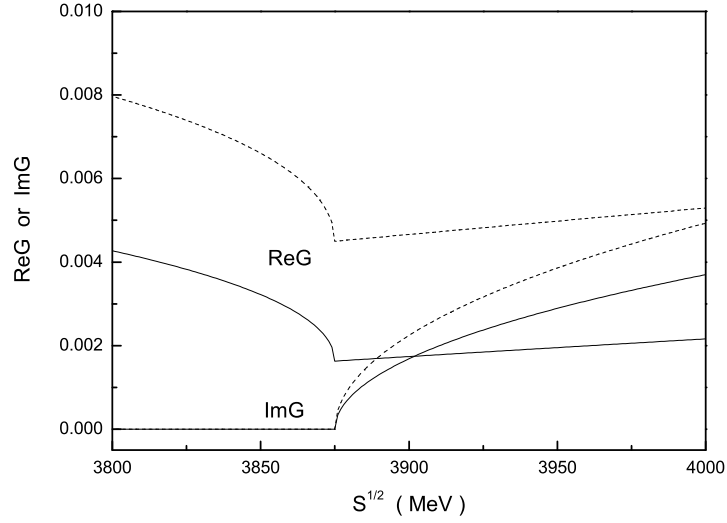}
}
\caption{The real and imaginary parts of the loop function of $D \bar{D}^*$ in Eq.~(\ref{eq:20170820}) .vs. the total energy $\sqrt{s}$ in the center of mass frame. The solid line denotes the case where the longitudinal propagator of the vector meson is taken into account, and the dash line stands for the case where only the transversal propagator of the vector meson is included.}
\label{fig:G_DD_mu1000}
\end{figure}

We found the results
are not sensitive to the values of $\lambda_1$, $\lambda_2$ and $\lambda_3$ in the
$\tilde{V}_{{D \bar{D}^* \rightarrow D \bar{D}^*}}$ in Eq.~(\ref{eq:tildeV1}) and $\tilde{V}_{{D \bar{D}^* \rightarrow \pi^0 \psi}}$ in Eq.~(\ref{eq:tildeV2}), so
We choose $\lambda_1=-1$, $\lambda_2=-\frac{1}{f_\pi}$ and $\lambda_3=\frac{1}{f_\pi}$ in the calculation.
When the Bethe-Salpeter equation is solved, the value of the subtraction constant in the loop function is fixed to be $a=-2$, while the regularization scale is chosen to be $\mu=500$MeV.
The real and imaginary parts of the $D \bar{D}^*$ loop function are depicted in Fig.~\ref{fig:G_DD_mu1000}, where the solid line denotes those of the loop function with the longitudinal part of the vector meson propagator taken into account, and the dash line stands for the original case that only the transversal part of the vector meson propagator is
included. It manifests that the real part of the loop function is only about half of the original values when the longitudinal part of the vector meson propagator is taken into account. Moreover, the imaginary part of the loop function above the threshold of $D \bar{D}^*$ is less than that of the original case.

%

\begin{figure}[!htb]
\centerline{
\includegraphics[width = 0.65\linewidth]{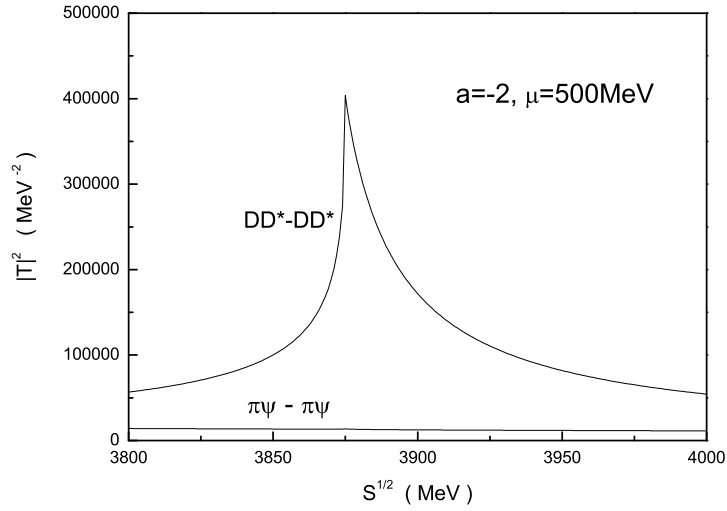}
}
\caption{The squared amplitudes as functions of the total energy $\sqrt{s}$ in the center of mass frame.
}
\label{fig:T2_mu1000}
\end{figure}

The squared amplitudes $|\tilde{T}_{ii}|^2$ as functions of the total energy $\sqrt{s}$ in the center of mass frame are depicted in Fig.~\ref{fig:T2_mu1000}, where the cases of $D \bar{D}^* \rightarrow D \bar{D}^*$ and $\pi \psi \rightarrow \pi \psi$ are labeled in the figure, respectively. A pole appears apparently in the region of 3900MeV. Actually, this resonance state is generated dynamically at the position of $3876-i9$MeV in the second Riemann sheet of the complex energy plane of $\sqrt{s}$, and can be associated to the $Zc(3900)$ particle consistently.
If the original form of the loop function in Eq.~(\ref{eq:Gpr}) is used in the calculation, a pole would appear at $3876-i33$MeV in the complex energy plane of $\sqrt{s}$,
It is apparent that the influence of the longitudinal part of the vector meson propagator is not important.

The couplings of this resonance state to $D \bar{D}^*$ and $\pi \psi$ are listed in Table~
\ref{table:coupings}. Apparently, the resonance state couples strongly to $D \bar{D}^*$.

\begin{table}[htbp]
\begin{tabular}{ccc}
\hline\hline
                   &    $g_i$     &    $|g_i|$   \\
\hline
 $D \bar{D}^*$     & $2.4+i1.0$ &    $2.6$  \\
 $\pi \psi$        & $0.1-i0.4$ &   $0.4$     \\
 \hline \hline
\end{tabular}
\caption{Couplings of the resonance state to $D \bar{D}^*$ and $\pi \psi$ in
the isospin $I=1$ sector.}
\label{table:coupings}
\end{table}

\section{Summary}
\label{sect:summary}

According to the effective Lagrangian of the $D$ meson, the $\bar{D}^*$ meson, the $J/\psi$ particle and the $\pi$ meson, the interaction between $D \bar{D}^*$ and $\pi \psi$ is studied in the unitary coupled-channel approximation.
The loop function of the vector meson and the pseudoscalar meson is calculated explicitly in the dimensional regularization scheme, and the longitudinal part of the vector meson propagator is taken into account.
Moreover, we think the polarization vectors of the initial and final vector mesons should be eliminated in the kernel of the vector meson and the pseudoscalar meson when the Bethe-Salpeter equation is solved. It means that all out-lines in the Feynman diagrams should be cut off and only the vertex of the vector meson- pseudoscalar meson interaction is necessary in the calculation. Furthermore, it is proved that the unitarity is not broken in the calculation.
In the isospin $I=1$ sector, a resonance state with a decay width about 20MeV is generated dynamically around 3900MeV.
This resonance state couples strongly to $D \bar{D}^*$, and it is assumed that this state is associated to the $Z_c(3900)$ particle in the PDG data.

\begin{acknowledgments}
We would like to thank Han-Qing Zheng, Jing-Long Pang and Qin-Rong Gong for useful discussions.
\end{acknowledgments}

\section*{Appendix}

In the Appendix part, we will give an exact formula of $H^{00}_{ab}(s)$ and $H^{11}_{ab}(s)$ in Eq.~(\ref{eq:20170820}).
We suppose
\be
\label{eq:appendix-0}
g^{\mu \nu} H^{00}_{ab}(P^2)+P^\mu P^\nu H^{11}_{ab}(P^2)=\frac{\mu^{4-d}}{i}
\int \frac{d^d k}{(2\pi)^d} \frac{k^\mu k^\nu}{(k^2-M_a^2+i\varepsilon)[(P-k)^2-M_b^2+i\varepsilon]},
\ee
with $P$ the total momentum of the system and $\mu$ the dimensional regularization scale.

In the $d-$dimension space, $g_{\mu\nu} g^{\mu \nu}=d$, and thus we can obtain
\ba
\label{eq:appendeix-1}
d H^{00}_{ab}(P^2)+P^2 H^{11}_{ab}(P^2)
=I_b+M_a^2 H_{ab}(P^2),
\ea
where
\be
I_b=\frac{\mu^{4-d}}{i} \int \frac{d^d k}{(2\pi)^d} \frac{1}{(k^2-M_b^2+i\varepsilon)}
=-\frac{M_b^2}{16\pi^2}
\left( R+\ln \frac{M_b^2}{\mu^2} \right),
\ee
%
with $R=a_l(\mu)+1$ and $a_l(\mu)$ the subtraction constant,
and
\be
H_{ab}(P^2)=\frac{\mu^{4-d}}{i}
\int \frac{d^d k}{(2\pi)^d} \frac{1}{(k^2-M_a^2+i\varepsilon)[(P-k)^2-M_b^2+i\varepsilon]}.
\ee
On the limit of $d \rightarrow 4$, $H_{ab}(P^2) \rightarrow -G_{ab}(s)$.

Similarly, multiply Eq.~(\ref{eq:appendix-0}) by $P^\mu P^\nu$ to obtain
\ba
\label{eq:appendeix-2}
P^2 H^{00}_{ab}(P^2)+P^4 H^{11}_{ab}(P^2)
=\frac{1}{2} [P^2 I_b - (P^2+\Delta_{ab})P^2 H^1_{ab}(P^2) ],
\ea
where
\be
-P^\mu H^1_{ab}(P^2)=\frac{\mu^{4-d}}{i}
\int \frac{d^d k}{(2\pi)^d} \frac{k^\mu}{(k^2-M_a^2+i\varepsilon)[(P-k)^2-M_b^2+i\varepsilon]},
\ee
and
\be
\label{eq:appendix-3}
H^1_{ab}(P^2)=\frac{1}{2P^2} \left[ I_a-I_b-(P^2+\Delta_{ab})H_{ab}(P^2) \right],
\ee
with $\Delta_{ab}=M_a^2-M_b^2$. The proof of Eq.~(\ref{eq:appendix-3})can be found in the appendix part of Ref.~\cite{Dong:2016auh}.

According to Eqs.~(\ref{eq:appendeix-1}) and (\ref{eq:appendeix-2}), we can obtain
\ba
\label{eq:H00}
H^{00}_{ab}(s)
&=&\frac{1}{12s} \{(s+\Delta_{ab})I_a +(s-\Delta_{ab})I_b +[4s M_a^2-(s+\Delta_{ab})^2]H_{ab}(s)
\}\nn \\&&-\frac{1}{16\pi^2} \frac{1}{18} (s-3\Sigma_{ab}),
\ea
and
\ba
\label{eq:H11}
H^{11}_{ab}(s)
&=&\frac{1}{3s^2} \{-(s+\Delta_{ab})I_a +(2s+\Delta_{ab})I_b -[s M_a^2-(s+\Delta_{ab})^2]H_{ab}(s)
\}\nn \\&&+\frac{1}{16\pi^2} \frac{1}{18s} (s-3\Sigma_{ab}),
\ea
with $s=P^2$ and $\Sigma_{ab}=M_a^2+M_b^2$.

\end{document}